\documentclass[12pt]{article}
\usepackage{amssymb}
\usepackage{epsfig}
\usepackage{amssymb,amsmath}
\usepackage{slashed}
\usepackage{multirow}

 \makeatletter
\def\alt{\mathrel{\mathpalette\gl@align<}}
\def\agt{\mathrel{\mathpalette\gl@align>}}
\def\gl@align#1#2{\lower.6ex\vbox{\baselineskip\z@skip\lineskip\z@
\ialign{$\m@th#1\hfil##\hfil$\crcr#2\crcr\sim\crcr}}} \makeatother

\begin{document}

\begin{center}
\baselineskip 20pt {\Large\bf 
Higgs Inflation, Quantum Smearing and the Tensor to Scalar Ratio}

\vspace{.5cm}

{\large Mansoor Ur Rehman and Qaisar Shafi} \vspace{.5cm}

{\baselineskip 20pt \it 
Bartol Research Institute, Department of Physics and Astronomy, \\
University of Delaware, Newark, DE 19716, USA \\
}

\vspace{1cm}
\end{center}

\begin{abstract}

In cosmic inflation driven by a scalar gauge singlet field with a tree level 
Higgs potential, the scalar to tensor ratio $r$ is estimated to be larger than 
$0.036$, provided the scalar spectral index $n_s \geq 0.96$. We discuss quantum 
smearing of these predictions arising from the inflaton couplings to other particles 
such as GUT scalars, and show that these corrections can significantly 
decrease $r$. However, for $n_s \geq 0.96$, we obtain $r \geq 0.02$ which can 
be tested by the Planck satellite.

\end{abstract}

\date{}


In any realistic inflationary cosmology \cite{Cosmology} 
the scalar inflaton field must
couple to additional fields in order to implement the transition to
a radiation dominated universe. These couplings, through quantum
corrections (smearing), modify the tree level inflationary
potential and therefore the corresponding predictions for the scalar
spectral index $n_s$, tensor to scalar ratio $r$ 
(measure of gravity waves), and the running of the spectral 
index  $dn_s/d \ln k$. 
In Ref. \cite{NeferSenoguz:2008nn}, it was shown 
that quantum corrections of the chaotic inflationary potential, 
computed ala Coleman-Weinberg \cite{Coleman:1973jx}, 
and induced via couplings of the inflaton to other
fields can significantly modify the tree level inflationary predictions. 
For instance, in the case of 
chaotic $\phi^2$ inflation \cite{Linde:1983gd} supplemented by 
a Yukawa interaction, the 
tree level prediction of $n_s = 0.966$ and $r = 0.13$ is replaced by 
$0.93 \lesssim n_s \lesssim 0.966$ and $0.023 \lesssim r \lesssim 0.135$ 
\cite{NeferSenoguz:2008nn}. A similar analysis has also been carried out 
for a non-supersymmetric hybrid inflationary model in Ref.~\cite{Rehman:2009wv}, 
where inclusion of these quantum corrections is shown to allow even sub-Planckian 
values of the inflaton field, consistent with the red tilted spectral index 
exhibited in the WMAP 7 year (WMAP7) results \cite{Komatsu:2010fb}.
The presence of a suitable Yukawa coupling, especially one involving right 
handed neutrinos, has one significant feature as far as realistic inflation model 
building is concerned. While enabling the transition from an inflaton to a 
radiation dominated universe, the decay products in this case contain right 
handed neutrinos whose subsequent out of equilibrium decay can give rise to 
the observed baryon asymmetry via leptogenesis (thermal \cite{Fukugita:1986hr} 
or non-thermal \cite{Lazarides:1991wu}). It was shown in 
\cite{NeferSenoguz:2008nn,Rehman:2009wv} that even though the radiative 
corrections during inflation may be sub-dominant, they can make 
sizeable corrections to the tree level predictions for $n_s$ and $r$. 
Therefore, it is interesting to analyze the effects of these radiative corrections
for various realistic inflationary models.

Motivated by the above observations, we investigate in this letter 
the impact of quantum corrections on the predictions of a
Higgs inflation model in which a gauge singlet scalar $\phi$ plays the 
role of the inflaton field. For a tree level treatment of this model, 
see Refs. \cite{Kallosh:2007wm,Smith:2008pf,Rehman:2008qs}, 
where it is shown that $\phi$ field has a trans-Planckian vacuum 
expectation value. Here we mainly consider
the two important interaction terms of $\phi$ in the 
renormalizable Lagrangian: 
a quartic interaction between $\phi$ and a GUT symmetry 
breaking scalar boson $\Phi$ ($\sim \lambda_{\Phi}\,\phi^2\,\Phi^2$),
and a Yukawa interaction between $\phi$ and a right handed Majorana neutrino 
$N$ ($\sim y\,\phi \,\bar{N}\,N$). 
It turns out that in order to obtain significant radiative corrections, both 
the $\Phi$ and $N$ fields should be heavier than inflaton, and hence 
do not significantly contribute to reheating. 
Therefore, we keep the right handed neutrinos
light enough to participate in reheating, while letting the GUT scalar boson
be heavy enough to produce considerable radiative corrections.
These radiative corrections are then shown to have a significant impact on 
the tree level predictions of $n_s$, $r$ and d$n_s/\ln k$. 
A very precise measurement of $n_s$ would be an extremely 
useful first step in the search for the correct inflation model. 
If $n_s$ can be determined very precisely, say by 
the Planck satellite experiment, the different predictions 
for $r$ which we report here would be an effective way to 
look for the quantum smearing effects that we have considered.
In practice, we find that for $n_s \geq 0.96$, the scalar 
to tensor ratio $r \geq 0.02$, which can be tested by the 
Planck satellite.


We consider the Lagrangian density \cite{NeferSenoguz:2008nn}
\begin{eqnarray}
{\cal L} &=&  \frac12\partial^{\mu}\phi_B\partial_{\mu}\phi_B 
+ \frac12\partial^{\mu}\Phi\partial_{\mu}\Phi 
+ \frac{i}{2}\bar{N}\gamma^{\mu}\partial_{\mu}N
  \nonumber \\ 
&& + \frac{1}{2}\,m_B^2\,\phi_B^2 - \frac{\lambda_B}{4}\,\phi_B^4 
-\frac{1}{2}y_B \phi_B \bar{N}N + \frac{1}{2}\lambda_{\Phi_B}^2\phi_B^2\Phi^2 
- \frac{a}{4}\Phi^4,\label{lagrangian}
\end{eqnarray}
where the subscript `B' denotes bare quantities, and $\Phi$ 
represents the GUT symmetry breaking scalar boson. The GUT 
symmetry is broken when $\Phi$ acquires a non-zero 
vacuum expectation value (VEV) 
$\langle \Phi \rangle = (\lambda_{\Phi}/a)^{1/2}\,\phi$. In order 
to keep the discussion simple, we have introduced a single right handed 
neutrino $N$ with Yukawa coupling $y_B$, and we ignore the bare mass
terms for $\Phi$ and $N$. In a more realistic 
scenario, successful leptogenesis requires at least two right-handed 
neutrinos.

The inflationary potential including one loop corrections 
\cite{Coleman:1973jx}, in terms of renormalized quantities, can be written as
\begin{eqnarray}
V &=& V_0 - \frac{1}{2}m^2\phi^2 + \left( \frac{\lambda}{4}+\frac{\lambda_{\Phi}}{4\,a} \right) \phi^4 
 + A\,\phi^4\,\left( \ln \left(\frac{\phi}{M} \right) + C \right),\label{}
\end{eqnarray}
where 
\begin{equation} \label{A}
A = \frac{1}{32\pi^2} \left( \mathcal{N} \lambda_{\Phi}^4 - 2 y^4 \right).
\end{equation}
We have assumed $A \gg \lambda^2$ and $A \gg (m / \phi)^4$ so that
the radiative correction from inflaton self couplings 
is suppressed \cite{NeferSenoguz:2008nn}. Also, 
with $m_{\Phi}^2 \approx \lambda_{\Phi}^2\,\phi^2 \gg H^2$ 
($H=$ Hubble constant), the `flat space' 
quantum correction is a good approximation during inflation. 
Here $\mathcal{N}$ is the
dimensionality of the representation of the field $\Phi$, $V_0$ is the vacuum energy
density at the origin and $C$ is a constant which can be determined from the minimization
condition $V'(M) = 0$, where $M = \langle \phi \rangle$ denotes the VEV of 
$\phi$ at the minimum. Furthermore, if we require the potential to be zero at $M$, i.e. 
$V(M) = 0$, the value of $V_0$ is determined and the potential then
takes the following form
\begin{eqnarray}
V = \left( \frac{m^2\,M^2}{4} \right)\,\left[ 1 - \left( \frac{\phi}{M}\right)^2 \right]^2
+A\,\phi^4\,\left[ \ln \left(\frac{\phi}{M}\right) -\frac{1}{4} \right]
+ \frac{A\,M^4}{4},
\end{eqnarray}
where $V(\phi = 0) \equiv V_0 = \frac{m^2\,M^2}{4}+\frac{A\,M^4}{4}$. 
The minimum of the above potential continues to lie at $M$ as long 
as $m^2 + 2 A M^2 > 0$ is satisfied. This condition is trivially met for
the $A>0$ case in which we are interested.
The first term in the above potential is the usual tree 
level Higgs potential, whereas the second term is the 
Coleman-Weinberg Potential (CWP), and embodies the radiative corrections.

For the sake of completeness we first review some of the salient features 
and predictions of tree level Higgs inflation. 
The tree level Higgs potential can be written 
as \cite{Kallosh:2007wm,Smith:2008pf,Rehman:2008qs}
\begin{equation}  \label{HP}
V_{\text{tree}} =  V_0\left[1-\left( \frac{\phi}{M}\right)^2\right]^2,
\end{equation}
with $V_0 = \frac{m^2\,M^2}{4}$. As discussed in Refs. \cite{Kallosh:2007wm,Rehman:2008qs}, inflation 
may occur above or below the VEV  $M$. For shorthand, we henceforth denote these regimes as the 
BV (below VEV) and AV (above VEV) solutions.

In the leading order approximation the 
slow-roll parameters are given as \cite{Cosmology}
\begin{equation}
\epsilon =
 \frac{m_P^2}{2}\left(\frac{V'}{V}\right)^2 \,,\quad
\eta = m_P^2\,\left(\frac{V''}{V}\right) \,,\quad
\xi^2 = m_P^4\,\left(\frac{V'\, V'''}{V^2}\right) \,,
\end{equation}
where $m_P \simeq 2.4 \times 10^{18}$ GeV is the reduced Planck
mass. The slow-roll approximation is valid as long as the conditions 
$\epsilon \ll 1$, $|\eta| \ll 1$ and $\xi^2 \ll 1$ hold. In this case 
the scalar spectral index $n_{s}$, the tensor-to-scalar ratio $r$, and the 
running of the spectral index $\frac{d n_{s}}{d \ln k}$ are given by
\begin{eqnarray}
n_{s} \!&\simeq&\! 1 - 6 \epsilon + 2 \eta, \label{ns}\\
r \!&\simeq&\! 16 \epsilon, \\
\frac{d n_{s}}{d \ln k} \!\!&\simeq&\!\!
16 \epsilon \eta - 24 \epsilon^2 - 2 \xi^2.
\end{eqnarray}
The number of e-foldings after the comoving scale $l_0=2\pi/k_0$ has crossed the horizon is
given by
\begin{equation} \label{efold1}
N_0=\frac{1}{2\,m_P^2}\int^{\phi_0}_{\phi_e}\frac{H(\phi)\rm{d}\phi}{H'(\phi)}, \end{equation}
where $\phi_0$ is the value of the field when the scale corresponding to 
$k_0$ exits the horizon, and $\phi_e$ is the value of the field at the 
end of inflation. The value of $\phi_e$ is given by the condition 
$2m_P^2(H'(\phi_e)/H(\phi_e))^2=1$, which can be calculated from 
the Hamilton-Jacobi equation \cite{Salopek:1990jq}
\begin{equation}
[H'(\phi)]^2-\frac{3}{2\,m_P^2}H^2(\phi)=-\frac{1}{2\,m_P^4}V(\phi)\,.
\end{equation}
Another expression of $N_0$ which explicitly depends on the 
thermal history of the universe is 
give by \cite{Kolb:1990vq,NeferSenoguz:2008nn}
\begin{equation} \label{nuk}
N_0\approx65 + 2\,\ln \left[\frac{V(\phi_0)^{1/4}}{m_P} \right]-\frac{4}{3\gamma_{reh}}\ln \left[\frac{V(\phi_e)^{1/4}}{m_P} \right]
+\left(\frac{4}{3\gamma_{reh}}-1\right)\ln \left[\frac{\rho_{reh}^{1/4}}{m_P} \right]
\,,
\end{equation}
where $\rho_{reh}$ is the energy density at the end of reheating, and
$\gamma_{reh} = 2\,n/(n+2)$ for $V\propto\phi^n$ \cite{Turner:1983he}.
In particular, for $\phi^2$ and $\phi^4$ inflation $\gamma_{reh} = 1$ 
and $\gamma_{reh} = 4/3$ respectively.
In the latter case $N_0$ is independent of $\rho_{reh}$. 
In our numerical calculations we take $\gamma_{reh} = 1$ because 
the quadratic component dominates near the minimum. 
Moreover, we assume 
quantum corrections to $\gamma_{reh}$ to be negligible \cite{NeferSenoguz:2008nn}.
We can write $\rho_{reh} = \frac{\pi^2}{30}\,g_*\,T_{reh}^4$,
where $T_{reh}$ is the reheat temperature given by \cite{Martin:2006rs}
\begin{equation}
T_{reh} \simeq \left[ \frac{30/\,g_*}{2\pi^3(1+w_{reh})(5-3w_{reh})}\right]^{1/4}\sqrt{\Gamma_{\phi}\,m_P}.
\end{equation}
Here $w_{reh} = \gamma_{reh} -1$ is the equation 
of state parameter for the dominant component during the reheating 
phase, $g_* = 106.75$ and $\Gamma_{\phi} \approx y^2\,m_{\phi}/(8\,\pi)$ is the
inflaton decay width. 


The amplitude of the curvature perturbation is given by
\begin{equation} \label{perturb}
\Delta_{\mathcal{R}}=\frac{1}{2\sqrt{3}\pi\,m_P^3}\frac{V^{3/2}}{|V'|}\,,
\end{equation}
where $\Delta_{\mathcal{R}} = 4.93\times 10^{-5}$ at $k_0 = 0.002\, \rm{Mpc}^{-1}$
according to WMAP7 data \cite{Komatsu:2010fb}. Note that, for added precision, we include 
in our calculations the first order corrections in the slow-roll expansion for 
the quantities $n_s$, $r$, $d n_s /d\ln k$, and $\Delta_{\mathcal{R}}$ \cite{Stewart:1993bc}.

\begin{table}[t]
{\centering
\resizebox{!}{3.0 cm}{
\begin{tabular}{||c|c|c|c|c|c|c|c|c|c|c||}
\hline
 $V^{1/4}_0$(GeV) & $V(\phi_0)^{1/4}$(GeV) &  $M/m_P$ & $\phi_0/m_P$ & $\phi_e/m_P$ & $n_s$ & $r$ 
 & $m_{\phi}$ (GeV)& $T_{reh}$(GeV) & $N_0$ &  $\frac{dn_s}{d \ln k}\,(10^{-4}) $   \\ 
 \hline \hline 
 1.3$\times 10^{16}$ & 1.3$\times 10^{16}$ & 12.7 & 2.03 & 11.7 & 0.944 & 0.020 & 1.5$\times 10^{13}$ &
   5.1$\times 10^7$ & 54.9 & -3.6  \\ \hline
 1.4$\times 10^{16}$ & 1.4$\times 10^{16}$ & 14.0 & 2.86 & 13.0 & 0.950 & 0.028 & 1.6$\times 10^{13}$ &
   5.3$\times 10^7$ & 55.0 & -4.3  \\ \hline
 1.5$\times 10^{16}$ & 1.4$\times 10^{16}$ & 14.8 & 3.42 & 13.8 & 0.953 & 0.033 & 1.7$\times 10^{13}$ &
   5.3$\times 10^7$ & 55.0 & -4.6  \\ \hline
 1.6$\times 10^{16}$ & 1.5$\times 10^{16}$ & 16.8 & 4.93 & 15.9 & 0.958 & 0.045 & 1.8$\times 10^{13}$ &
   5.0$\times 10^7$ & 55.1 & -5.2  \\ \hline
 1.7$\times 10^{16}$ & 1.6$\times 10^{16}$ & 18.1 & 5.94 & 17.1 & 0.960 & 0.051 & 1.8$\times 10^{13}$ &
   4.8$\times 10^7$ & 55.2 & -5.4  \\ \hline
 1.8$\times 10^{16}$ & 1.6$\times 10^{16}$ & 19.5 & 7.15 & 18.6 & 0.961 & 0.058 & 1.9$\times 10^{13}$ &
   4.6$\times 10^7$ & 55.2 & -5.7  \\ \hline
 1.9$\times 10^{16}$ & 1.7$\times 10^{16}$ & 21.2 & 8.61 & 20.3 & 0.962 & 0.065 & 1.9$\times 10^{13}$ &
   4.2$\times 10^7$ & 55.2 & -5.9  \\ \hline
 1.9$\times 10^{16}$ & 1.7$\times 10^{16}$ & 23.2 & 10.3 & 22.2 & 0.963 & 0.072 & 1.9$\times 10^{13}$ &
   3.9$\times 10^7$ & 55.2 & -6.0 \\ \hline
 2.0$\times 10^{16}$ & 1.8$\times 10^{16}$ & 25.4 & 12.3 & 24.4 & 0.964 & 0.078 & 1.9$\times 10^{13}$ &
   3.6$\times 10^7$ & 55.2 & -6.1  \\ \hline
 2.4$\times 10^{16}$ & 1.9$\times 10^{16}$ & 37.2 & 23.6 & 36.2 & 0.965 & 0.101 & 1.9$\times 10^{13}$ &
   2.4$\times 10^7$ & 55.2 & -6.4  \\ \hline
 2.8$\times 10^{16}$ & 1.9$\times 10^{16}$ & 50.0 & 36.0 & 49.0 & 0.965 & 0.112 & 1.8$\times 10^{13}$ &
   1.7$\times 10^7$ & 55.2 & -6.5  \\ \hline
 3.2$\times 10^{16}$ & 2.0$\times 10^{16}$ & 67.0 & 52.8 & 66.0 & 0.965 & 0.121 & 1.8$\times 10^{13}$ &
   1.3$\times 10^7$ & 55.1 & -6.6  \\ \hline
 3.7$\times 10^{16}$ & 2.0$\times 10^{16}$ & 89.5 & 75.2 & 88.5 & 0.964 & 0.127 & 1.8$\times 10^{13}$ &
   9.2$\times 10^6$ & 55.1 & -6.6  \\ \hline
\end{tabular} 
\par} 
\caption{Predicted values of various inflationary parameters using the tree level Higgs potential.
Here we show only those values which fall inside the WMAP7 1$\sigma$ bounds (see Fig.~1).} \label{tableI}
\centering}
\end{table}

\begin{figure}[t] 
\begin{center} 
\includegraphics[width = 12cm]{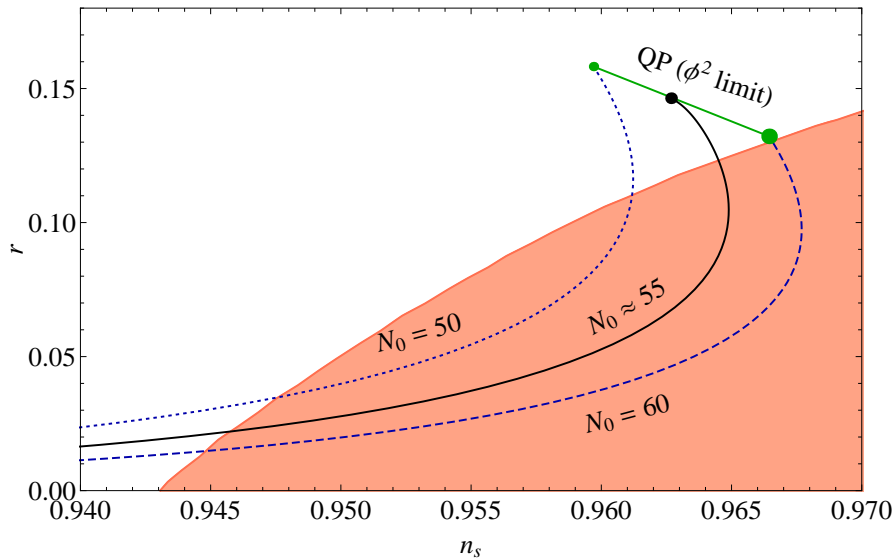}
\caption{$r$ vs. $n_s$ for tree level Higgs inflation. The blue
dotted, blue dashed and black solid curves correspond to 
number of e-foldings $N_0 = 50$,
$N_0 = 60$ and $N_0$ given in Eq.~(\ref{nuk}), respectively.
Small (big) green circles correspond to the Quadratic Potential (QP) 
with $N_0 = 50$ ($N_0 = 60$).} \label{TLHP}
\end{center}
\end{figure}

The predicted values of various parameters of the tree level Higgs inflation
can be obtained by using Eqs.~(\ref{HP}-\ref{perturb}) above, and 
these are displayed in Table \ref{tableI}. As we see,
the quantities $M$, $\phi_0$ and $\phi_e$ carry trans-Planckian values but 
still the vacuum energy scale during observable inflation is 
well below $m_P$. This implies that the quantum gravity effects
are relatively unimportant here. Moreover, very low values of the reheat temperature 
$T_{reh} \sim 10^7$ GeV $\ll m_{\phi} \sim 10^{13}$ GeV are 
realized with the Yukawa coupling $y \sim 10^{-6}$.  
Since we only consider the right handed Majorana neutrinos for 
reheating, we obtain negligible radiative corrections with 
their masses $m_N \leq m_{\phi}/2$ (or the Yukawa coupling 
$y \lesssim 10^{-6}$).
Taking the number of e-foldings $N_0$ between 50 to 60,
we obtain $0.945 \lesssim n_s \lesssim 0.967$ and 
$0.02 \lesssim r \lesssim 0.13$ within the 
WMAP7 1$\sigma$ bounds as shown in Fig.~\ref{TLHP}.
Furthermore, we only show the BV branch in Fig.~\ref{TLHP}
which lies largely within the WMAP7 1$\sigma$ bounds, 
whereas, the AV branch remains outside of the WMAP7 1$\sigma$ bounds
\cite{Rehman:2008qs} at tree level. In this letter we mainly restrict 
our discussion to the WMAP7 1$\sigma$ bounds.

The various limits of tree level Higgs inflation has been discussed in 
Ref. \cite{Rehman:2008qs}. For example, for $\phi \ll M$, the 
BV branch approaches the effective potential of new inflation 
$V \approx V_0\left( 1-2\left(\frac{\phi}{M} \right)^2 \right)$ 
in a region disfavored by WMAP7. On the other hand, 
for $\phi \gg M$, the AV branch approaches the effective 
potential of quartic inflation 
$V \approx \left(\frac{V_0}{M^4} \right)\phi^4$. 
As mentioned above this AV branch lies outside of the 
WMAP7 1-$\sigma$ bounds.
Finally, near the VEV we obtain an 
effective potential of quadratic inflation 
$V \approx \frac{1}{2} m_{\phi}^2 (\Delta\phi)^2$, where 
$\Delta\phi=M-\phi$ denotes the deviation of the field from the minimum and 
$m_{\phi}=\frac{2\sqrt{2V_{0}}}{M}$ is the inflaton mass. 
This chaotic inflationary model \cite{Linde:1983gd} predicts $m_{\phi}\simeq2\times10^{13}$ GeV, $\Delta\phi_0\simeq2\sqrt{N_0}$, $n_s\simeq 1-\frac{2}{N_0}$, and $r\simeq 4(1-n_s)$, corresponding to $V(\phi_0)\simeq(2\times10^{16}$ GeV$)^4$.  In fact this is the region in which the two branches 
meet, i.e. both the BV and AV branches converge to quadratic inflation in the high-$V_0$ limit.

Let us now investigate the effect of including quantum smearing to the 
tree level Higgs inflation.
In general we may consider two types of radiative corrections: fermionic 
radiative corrections (which arise from the Yukawa interaction that couples
$\phi$ to fermions) and bosonic radiative corrections 
(which arise from the quartic interaction between $\phi$ and other bosons). 
As we see in Eq.~(\ref{A}), these 
radiative corrections appear in the potential with opposite signs and 
also exhibit different behavior in various inflationary predictions.
These corrections become important around $|A| \sim 10^{-14}$ and if we 
consider the bosonic radiative corrections only, the mass of the  gauge 
scalar boson turns out to be of order the GUT scale 
$M_G \simeq  \lambda_{\Phi}\langle \phi \rangle \sim 10^{16}$ GeV, with
$\langle \phi \rangle \sim M_P$ and $\lambda_{\Phi} \sim 10^{-3}$. Here 
$M_P = 1.2 \times 10^{19}$ GeV is the Planck mass. 
Similarly, in order to obtain sizable effects with the fermionic 
radiative corrections we need a Yukawa coupling $y$ of order $10^{-3}$.
This, in turn, makes the fermion masses superheavy $\sim 10^{16}$~GeV, 
larger than the inflaton mass $m_{\phi} \sim 10^{13}$ GeV. Thus, 
these fermions are unable to participate in the reheating process
and are also too heavy for seesaw physics.
Consequently, we utilize the right handed Majorana neutrinos for 
reheating, which generates negligible radiative corrections 
with their masses $m_N \leq m_{\phi}/2$ 
and the Yukawa coupling $y \lesssim 10^{-6}$.
In this regard, the Higgs inflationary model is different from the 
chaotic inflationary model where the same Yukawa interaction, 
responsible for the reheating,
has also been shown to make sizable contributions, 
through radiative corrections, in modifications of the
tree level predictions \cite{NeferSenoguz:2008nn}. 
This is related to the fact that the Yukawa couplings 
in the chaotic inflation are not restricted 
by a non-zero inflaton VEV. In the following discussion, 
therefore, we only consider the effect of bosonic radiative corrections 
from the GUT field $\Phi$.

\begin{figure}[t] 
\begin{center} 
\includegraphics[width = 12cm]{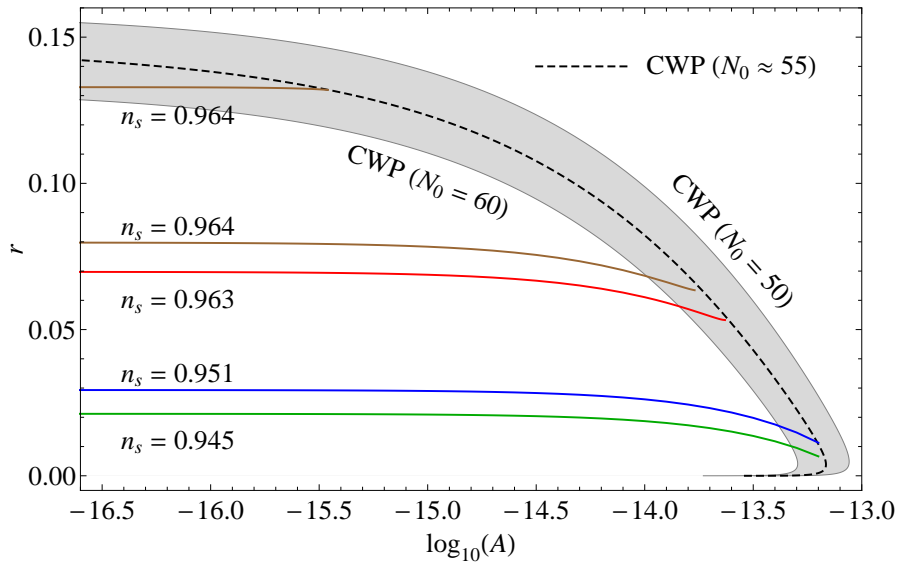} 
\caption{$r$ vs. log$_{10}(A)$ for the 
radiatively corrected Higgs potential for different 
values of $n_s$. Here the black dashed curve represents the predictions of
Coleman-Weinberg Potential (CWP) with the number of e-foldings $N_0$
given in Eq.~(\ref{nuk}).} \label{rA}
\end{center}
\end{figure}

\begin{figure}[t] 
\begin{center} 
\includegraphics[width = 12cm]{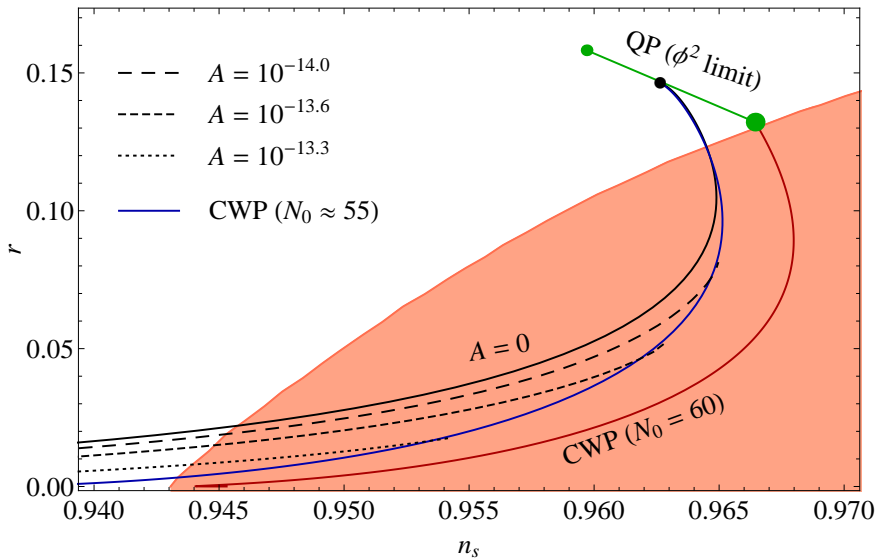} 
\caption{$r$ vs. $n_s$ for the tree level Higgs potential ($A = 0$)
and the radiatively corrected Higgs potential 
($A = 10^{-14.0}$, $10^{-13.6}$, $10^{-13.3}$).
Blue and red curves represent the predictions of Coleman-Weinberg 
Potential (CWP) with the number of e-foldings $N_0 \approx 55$
using Eq.~(\ref{nuk}), and $N_0 = 60$ respectively.
Small (big) green circles correspond to Quadratic Potential (QP) 
with $N_0 = 50$ ($N_0 = 60$).} \label{rns2}
\end{center}
\end{figure}

The predictions of the bosonic radiatively corrected Higgs potential
are shown in Fig.~\ref{rA}. The parameter $A$ here quantifies the 
amount of radiative corrections.
In order to explore the entire space of smearing, we show 
in Fig.~\ref{rA} a plot between $r$ and 
log$_{10}(A)$ for the values of $n_s$ which lie
within the WMAP7 1$\sigma$ bounds.
As we increase the value of $A$, the tensor to scalar ratio $r$
is observed to decrease. This is contrary to what we usually 
expect in the chaotic inflationary models where bosonic (fermionic) 
radiative corrections increase (reduce) the value of $r$. 
Actually, the BV branch of the tree level Higgs potential 
and the added bosonic radiative 
corrections have slopes with opposite signs, which reduces the magnitude of 
the total slope of the potential and in turn decreases the value 
of $r$. In contrast, both the tree level chaotic inflationary 
potentials and their bosonic radiative corrections share slopes with 
same signs and this raises the value of $r$. 
A similar reduction in $r$ has also been observed in 
Ref. \cite{Destri:2009wn} by adding higher order even 
powers of $\phi$ to the tree level Higgs potential. 
This reduction is prominent (insignificant) for small (large) 
values of the spectral index $n_s$ and the tensor 
to scalar ratio $r$ as a result of an increase (decrease) in the
value of $A$.
For example, with $n_s = 0.954$, $r \approx 0.02$ 
$(n_s = 0.964$, $r \approx 0.13)$ the reduction in 
$r$ is $\Delta r \approx 0.02$ ($\Delta r \approx 0$ ).
Furthermore, the CWP curve provides an upper bound on $A$ for a 
given value of $n_s$, as is clear from Fig.~\ref{rA}.
For instance, the upper bound on $A$ is $\sim 10^{-13.6}$ for 
the WMAP7 central value of spectral index $n_s = 0.963$. 
In the small $A$ limit, all curves reproduce the 
standard tree level Higgs inflation predictions. 

In Fig.~\ref{rns2}, we fix the value of $A$
and vary $m$ between zero and its tree level value. We consider, 
as an example, three different values of the parameter 
$A = 10^{-14.0}$, $A = 10^{-13.6}$, 
and $A = 10^{-13.3}$. Each curve of radiatively corrected
Higgs potential with constant value of $A$ 
approaches the CWP prediction in the small $m$ limit. 
As is clear from Fig.~\ref{rns2}, the CWP curve
is more contiguous to the central region of 
WMAP7 data in comparison to the tree level Higgs 
potential curve. Therefore, in Fig.~\ref{rns2} 
we also include a curve of CWP with $N_0 = 60$ 
which further extends the allowed
space towards the central region of 1$\sigma$ WMAP7 data.
Moreover, both the tensor to scalar ratio $r$ and the spectral 
index $n_s$ are observed to decrease with an increase in $A$
as explained above.
The predictions of radiatively corrected Higgs potential, in fact, 
interpolate between the tree level Higgs potential and 
the CWP curves. 

To see the extent of quantum spread, we provide in Table~\ref{tableIII} 
the predicted values of various parameters of this model 
corresponding to the the WMAP7 central value of the spectral index, 
$n_s = 0.963$. Most of the quantities exhibit a small variation
from their respective tree level predictions. However, the 
the tensor to scalar ratio $r$ experiences a significant shift.
The tree level prediction of $r = 0.07$ acquires a total quantum spread 
of $0.05 \lesssim r \lesssim 0.07$ with $0 \lesssim A \lesssim 10^{-13.6}$.

The CWP prediction represents the maximal smearing of the 
radiatively corrected Higgs inflation results. 
This model has been studied extensively in the past 
\cite{Shafi:1983bd,Pi:1984pv,Lazarides:1984pq} 
and recently \cite{Smith:2008pf,Rehman:2008qs,Shafi:2006cs}. 
The predicted values of various parameters of this model are 
displayed in Table \ref{tableII} with the number of e-foldings $N_0 = 60$.
Similar to tree level 
Higgs inflation, $M$, $\phi_0$ and $\phi_e$ carry trans-Planckian 
values with vacuum energy scale well below $m_P$ during 
observable inflation. We obtain $r \approx 0.03$ and $A \approx 3 \times 10^{-14}$
for the WMAP central value of the spectral index $n_s = 0.963$.

\begin{table}[t]
{\centering
\resizebox{!}{3.5 cm}{
\begin{tabular}{||c|c|c|c|c|c|c|c|c|c||}
\hline
 $V^{1/4}_0$(GeV) & $V(\phi_0)^{1/4}$(GeV) & $A$ &  $M/m_P$ & $\phi_0/m_P$ & $\phi_e/m_P$ & $r$ 
 & $m_{\phi}$(GeV) & $T_r$(GeV) & $\frac{dn_s}{d \ln k}\,(10^{-4}) $   \\ 
\hline \hline
 1.91$\times 10^{16}$ & 1.72$\times 10^{16}$ & $10^{-17.0}$ & 22.6 & 9.79 & 21.6  & 0.070 & 1.9$\times 
   10^{13}$ & 4.0$\times 10^7$  & -6.0 \\ \hline
 1.91$\times 10^{16}$ & 1.72$\times 10^{16}$ & $10^{-16.0}$ & 22.5 & 9.78 & 21.6  & 0.0670 & 1.9$\times 
   10^{13}$ & 4.0$\times 10^7$  & -6.0 \\  \hline
 1.90$\times 10^{16}$ & 1.72$\times 10^{16}$ & $10^{-15.0}$ & 22.4 & 9.72 & 21.5  & 0.069 & 1.9$\times 
   10^{13}$ & 4.1$\times 10^7$  & -6.0 \\  \hline
 1.90$\times 10^{16}$ & 1.71$\times 10^{16}$ & $10^{-14.8}$ & 22.4 & 9.67 & 21.4  & 0.068 & 1.9$\times 
   10^{13}$ & 4.1$\times 10^7$  & -6.0 \\  \hline
 1.89$\times 10^{16}$ & 1.71$\times 10^{16}$ & $10^{-14.6}$ & 22.3 & 9.62 & 21.3  & 0.067 & 1.9$\times 
   10^{13}$ & 4.1$\times 10^7$  & -6.0 \\  \hline
 1.87$\times 10^{16}$ & 1.70$\times 10^{16}$ & $10^{-14.4}$ & 22.1 & 9.53 & 21.2  & 0.066 & 1.9$\times 
   10^{13}$ & 4.2$\times 10^7$  & -6.0 \\  \hline
 1.85$\times 10^{16}$ & 1.69$\times 10^{16}$ & $10^{-14.2}$ & 22.0 & 9.44 & 21.0  & 0.064 & 1.9$\times 
   10^{13}$ & 4.2$\times 10^7$  & -6.0 \\  \hline
 1.82$\times 10^{16}$ & 1.67$\times 10^{16}$ & $10^{-14.0}$ & 21.7 & 9.36 & 20.8  & 0.061 & 1.9$\times 
   10^{13}$ & 4.3$\times 10^7$  & -6.0 \\  \hline
 1.82$\times 10^{16}$ & 1.66$\times 10^{16}$ & $10^{-13.9}$ & 21.7 & 9.35 & 20.7  & 0.060 & 1.9$\times 
   10^{13}$ & 4.3$\times 10^7$  & -6.0 \\  \hline
 1.80$\times 10^{16}$ & 1.65$\times 10^{16}$ & $10^{-13.8}$ & 21.6 & 9.40 & 20.7  & 0.058 & 1.9$\times 
   10^{13}$ & 4.4$\times 10^7$  & -6.0 \\  \hline
 1.78$\times 10^{16}$ & 1.64$\times 10^{16}$ & $10^{-13.8}$ & 21.7 & 9.48 & 20.7  & 0.057 & 1.9$\times 
   10^{13}$ & 4.4$\times 10^7$  & -6.0 \\  \hline
 1.76$\times 10^{16}$ & 1.62$\times 10^{16}$ & $10^{-13.7}$ & 22.0 & 9.91 & 21.0  & 0.055 & 2.0$\times 
   10^{13}$ & 4.4$\times 10^7$  & -6.1 \\  \hline
 1.75$\times 10^{16}$ & 1.61$\times 10^{16}$ & $10^{-13.7}$ & 22.4 & 10.4 & 21.5  & 0.054 & 2.0$\times 
   10^{13}$ & 4.3$\times 10^7$  & -6.1 \\  \hline
 1.75$\times 10^{16}$ & 1.61$\times 10^{16}$ & $10^{-13.6}$ & 23.3 & 11.3 & 22.3  & 0.053 & 2.0$\times 
   10^{13}$ & 4.1$\times 10^7$  & -6.1  \\  \hline
\end{tabular} 
\par} 
\caption{Predicted values of various inflationary parameters using 
the radiatively corrected Higgs potential with $n_s = 0.963$ 
and $N_0$ ($\approx 55$) given in Eq.~(\ref{nuk}).} \label{tableIII}
\centering}
\end{table}

\begin{table}[t]
{\centering
\resizebox{!}{5.9 cm}
{
\begin{tabular}{||c|c|c|c|c|c|c|c|c|c|c||}
\hline
 $V^{1/4}_0$(GeV) & $V(\phi_0)^{1/4}$(GeV) & $A$ &  $M/m_P$ & $\phi_0/m_P$ & $\phi_e/m_P$ & $n_s$ & $r$ 
 & $m_{\phi}$ (GeV)&  $\frac{dn_s}{d \ln k}\,(10^{-4})$    \\ 
 \hline \hline 
 1.3$\times 10^{16}$ & 1.3$\times 10^{16}$ & 3.9$\times 10^{-14}$ & 17.0
   & 6.24 & 16.1 & 0.960 & 0.020 & 1.6$\times 10^{13}$  & -5.3 \\ \hline 
 1.4$\times 10^{16}$ & 1.3$\times 10^{16}$ & 3.6$\times 10^{-14}$ & 18.1
   & 7.10 & 17.2 & 0.961 & 0.024 & 1.7$\times 10^{13}$ &  -5.2 \\ \hline 
 1.4$\times 10^{16}$ & 1.4$\times 10^{16}$ & 3.2$\times 10^{-14}$ & 19.4
   & 8.10 & 18.5 & 0.962 & 0.028 & 1.7$\times 10^{13}$ &  -5.1 \\ \hline      
 1.5$\times 10^{16}$ & 1.4$\times 10^{16}$ & 3.1$\times 10^{-14}$ & 20.2
   & 8.67 & 19.2 & 0.963 & 0.030 & 1.7$\times 10^{13}$ &  -5.1 \\ \hline 
 1.5$\times 10^{16}$ & 1.4$\times 10^{16}$ & 2.7$\times 10^{-14}$ & 21.7
   & 9.95 & 20.8 & 0.964 & 0.035 & 1.8$\times 10^{13}$ &  -5.1 \\ \hline 
 1.6$\times 10^{16}$ & 1.5$\times 10^{16}$ & 2.4$\times 10^{-14}$ & 23.5
   & 11.4 & 22.6 & 0.965 & 0.040 & 1.8$\times 10^{13}$ &  -5.1 \\ \hline 
 1.6$\times 10^{16}$ & 1.5$\times 10^{16}$ & 2.2$\times 10^{-14}$ & 24.5
   & 12.3 & 23.6 & 0.965 & 0.043 & 1.8$\times 10^{13}$ &  -5.1 \\ \hline 
 1.7$\times 10^{16}$ & 1.5$\times 10^{16}$ & 2.1$\times 10^{-14}$ & 25.3
   & 13.0 & 24.4 & 0.965 & 0.045 & 1.8$\times 10^{13}$ &  -5.1 \\ \hline 
 1.7$\times 10^{16}$ & 1.6$\times 10^{16}$ & 2.0$\times 10^{-14}$ & 26.1
   & 13.7 & 25.2 & 0.966 & 0.047 & 1.8$\times 10^{13}$ &  -5.1 \\ \hline 
 1.7$\times 10^{16}$ & 1.6$\times 10^{16}$ & 1.9$\times 10^{-14}$ & 26.7
   & 14.2 & 25.7 & 0.966 & 0.049 & 1.8$\times 10^{13}$ &  -5.1 \\ \hline 
 1.7$\times 10^{16}$ & 1.6$\times 10^{16}$ & 1.8$\times 10^{-14}$ & 27.3
   & 14.7 & 26.3 & 0.966 & 0.050 & 1.8$\times 10^{13}$ &  -5.1 \\ \hline    
 1.8$\times 10^{16}$ & 1.6$\times 10^{16}$ & 1.7$\times 10^{-14}$ & 28.2
   & 15.5 & 27.3 & 0.966 & 0.052 & 1.8$\times 10^{13}$ &  -5.1 \\ \hline 
 1.8$\times 10^{16}$ & 1.6$\times 10^{16}$ & 1.7$\times 10^{-14}$ & 28.5
   & 15.8 & 27.6 & 0.966 & 0.053 & 1.8$\times 10^{13}$ &  -5.1 \\ \hline 
 1.8$\times 10^{16}$ & 1.6$\times 10^{16}$ & 1.6$\times 10^{-14}$ & 28.9 &
   16.1 & 27.9 & 0.966 & 0.054 & 1.8$\times 10^{13}$   &  -5.2 \\ \hline 
 1.8$\times 10^{16}$ & 1.6$\times 10^{16}$ & 1.5$\times 10^{-14}$ & 29.9
   & 17.0 & 28.9 & 0.967 & 0.056 & 1.8$\times 10^{13}$ &  -5.2 \\ \hline 
 1.9$\times 10^{16}$ & 1.7$\times 10^{16}$ & 1.3$\times 10^{-14}$ & 32.0
   & 19.0 & 31.0 & 0.967 & 0.060 & 1.8$\times 10^{13}$ &  -5.2 \\ \hline 
 2.0$\times 10^{16}$ & 1.7$\times 10^{16}$ & 9.5$\times 10^{-15}$ & 37.8 &
   24.4 & 36.9 & 0.968 & 0.070 & 1.8$\times 10^{13}$ &  -5.3 \\ \hline 
 2.2$\times 10^{16}$ & 1.8$\times 10^{16}$ & 6.3$\times 10^{-15}$ & 45.5 &
   31.7 & 44.5 & 0.968 & 0.080 & 1.8$\times 10^{13}$ &  -5.4 \\ \hline 
 2.6$\times 10^{16}$ & 1.9$\times 10^{16}$ & 3.2$\times 10^{-15}$ & 62.2 &
   48.0 & 61.2 & 0.968 & 0.094 & 1.7$\times 10^{13}$ &  -5.5 \\ \hline 
 2.8$\times 10^{16}$ & 1.9$\times 10^{16}$ & 2.2$\times 10^{-15}$ & 74.1 &
   59.7 & 73.1 & 0.968 & 0.100 & 1.7$\times 10^{13}$ &  -5.5 \\ \hline 
 3.0$\times 10^{16}$ & 1.9$\times 10^{16}$ & 1.5$\times 10^{-15}$ & 90.4 &
   75.8 & 89.4 & 0.968 & 0.106 & 1.7$\times 10^{13}$ &  -5.5 \\ \hline 
 3.5$\times 10^{16}$ & 1.9$\times 10^{16}$ & 7.4$\times 10^{-16}$ & 124 &
   110 & 123 & 0.967 & 0.113 & 1.6$\times 10^{13}$ &  -5.6 \\ \hline 
 4.0$\times 10^{16}$ & 2.0$\times 10^{16}$ & 4.4$\times 10^{-16}$ & 159 &
   144 & 158 & 0.967 & 0.117 & 1.6$\times 10^{13}$ & -5.6 \\ \hline 
 4.5$\times 10^{16}$ & 2.0$\times 10^{16}$ & 2.4$\times 10^{-16}$ & 212 &
   197 & 211 & 0.967 & 0.121 & 1.6$\times 10^{13}$ &  -5.6 \\ \hline 
 5.0$\times 10^{16}$ & 2.0$\times 10^{16}$ & 1.6$\times 10^{-16}$ & 256 &
   241 & 255 & 0.967 & 0.123 & 1.6$\times 10^{13}$ &  -5.6 \\ \hline 
 6.0$\times 10^{16}$ & 2.0$\times 10^{16}$ & 7.5$\times 10^{-17}$ & 374 &
   358 & 373 & 0.967 & 0.126 & 1.6$\times 10^{13}$ &  -5.6 \\ \hline 
 7.0$\times 10^{16}$ & 2.0$\times 10^{16}$ & 3.9$\times 10^{-17}$ & 519 &
   504 & 518 & 0.967 & 0.128 & 1.6$\times 10^{13}$ &  -5.6 \\ \hline 
 7.9$\times 10^{16}$ & 2.0$\times 10^{16}$ & 2.4$\times 10^{-17}$ & 655 &
   640 & 654 & 0.967 & 0.129 & 1.6$\times 10^{13}$ &  -5.6 \\ \hline 
\end{tabular} 
\par} 
\caption{Predicted values of various inflationary parameters using 
the Coleman-Weinberg Potential (CWP).
Here we show only those values which fall inside the WMAP7 1$\sigma$ bounds
and satisfy $n_s \gtrsim 0.96$ and $N_0 = 60$.} \label{tableII}
\centering}
\end{table}

To summarize, we discuss the effect of including 
quantum corrections (smearing) in the tree level
Higgs inflation. In contrast to chaotic inflation, a
Yukawa interaction between inflaton and right handed neutrinos 
generates negligible radiative corrections owing to  
light ($\ll M_G$) neutrino masses. On the other hand, the bosonic radiative 
corrections, which are obtained from a quartic 
interaction between inflaton and a GUT symmetry 
breaking scalar boson, can yield noticeable effects. 
As a result of including these corrections, 
a reduction in the tensor to scalar ratio $r$ is observed.
The Coleman-Weinberg potential provides, in this case, the 
maximal quantum smearing limit.
Using the WMAP7 central value of the spectral 
index $n_s = 0.963$, the tree level prediction of 
$r \approx 0.07$ is replaced by $0.05 \lesssim r \lesssim 0.07$.
We emphasize that while working with high precision 
observations such as the current Planck satellite experiment 
we cannot ignore these radiative corrections in analyzing 
the predictions of various inflationary models.

\section*{Acknowledgments}
Q.S. thanks Steve Weinberg for a helpful correspondence 
regarding inflationary potentials and radiative corrections. 
We also thank Nefer {\c S}eno$\breve{\textrm{g}}$uz and Joshua R. Wickman 
for valuable discussions.
This work is supported in part by the DOE under grant 
No.~DE-FG02-91ER40626 (Q.S. and M.R.), and by the University of 
Delaware competitive fellowship (M.R.).


\begin{thebibliography}{99}


\bibitem{Cosmology}
   For reviews see  
   A. Linde, {\it Particle Physics and Inflationary Cosmology}
(Harwood Academic Publishers, 1990);
  D.~H.~Lyth and A.~Riotto,
  Phys.\ Rept.\  {\bf 314}, 1 (1999)
  [arXiv:hep-ph/9807278];
  A.~Mazumdar and J.~Rocher,
  arXiv:1001.0993 [hep-ph].



\bibitem{NeferSenoguz:2008nn}
  V.~N.~Senoguz and Q.~Shafi,
  Phys.\ Lett.\  B {\bf 668}, 6 (2008)
  [arXiv:0806.2798 [hep-ph]].


\bibitem{Coleman:1973jx}
  S.~R.~Coleman and E.~J.~Weinberg,
  Phys.\ Rev.\  D {\bf 7}, 1888 (1973);
  S.~Weinberg,
  {\it The Quantum Theory of Fields}. Vol. 2: Modern Applications, 
  (Cambridge University Press, 1996).  
  For a review and additional references, see
  M.~Sher,
  Phys.\ Rept.\  {\bf 179}, 273 (1989).
  
  
  
  
\bibitem{Linde:1983gd}
  A.~D.~Linde,
  Phys.\ Lett.\  B {\bf 129}, 177 (1983).


\bibitem{Rehman:2009wv}
  M.~U.~Rehman, Q.~Shafi and J.~R.~Wickman,
  Phys.\ Rev.\  D {\bf 79}, 103503 (2009)
  [arXiv:0901.4345 [hep-ph]].


\bibitem{Komatsu:2010fb}
  E.~Komatsu {\it et al.},
  arXiv:1001.4538 [astro-ph.CO].


\bibitem{Fukugita:1986hr}
  M.~Fukugita and T.~Yanagida,
  Phys.\ Lett.\  B {\bf 174}, 45 (1986).

\bibitem{Lazarides:1991wu}
  G.~Lazarides and Q.~Shafi,
  Phys.\ Lett.\  B {\bf 258}, 305 (1991).




\bibitem{Kallosh:2007wm}
  R.~Kallosh and A.~Linde,
  JCAP {\bf 0704}, 017 (2007)
  [arXiv:0704.0647 [hep-th]].



\bibitem{Smith:2008pf}
  T.~L.~Smith, M.~Kamionkowski and A.~Cooray,
  Phys.\ Rev.\  D {\bf 78}, 083525 (2008)
  [arXiv:0802.1530 [astro-ph]].


\bibitem{Rehman:2008qs}
  M.~U.~Rehman, Q.~Shafi and J.~R.~Wickman,
  Phys.\ Rev.\  D {\bf 78}, 123516 (2008)
  [arXiv:0810.3625 [hep-ph]].




\bibitem{Salopek:1990jq}
  D.~S.~Salopek and J.~R.~Bond,
  Phys.\ Rev.\  D {\bf 42}, 3936 (1990).


  
  
\bibitem{Kolb:1990vq}
  E.~W.~Kolb and M.~S.~Turner,
  Front.\ Phys.\  {\bf 69}, 1 (1990);
  A.~R.~Liddle and S.~M.~Leach,
  Phys.\ Rev.\  D {\bf 68}, 103503 (2003)
  [arXiv:astro-ph/0305263].

\bibitem{Turner:1983he}
  M.~S.~Turner,
  Phys.\ Rev.\  D {\bf 28}, 1243 (1983).


\bibitem{Martin:2006rs}
  J.~Martin and C.~Ringeval,
  JCAP {\bf 0608}, 009 (2006)
  [arXiv:astro-ph/0605367];
  E.W. Kolb and M.S. Turner, {\it The Early Universe} (Westview,
1990).

\bibitem{Stewart:1993bc}
  E.~D.~Stewart and D.~H.~Lyth,
  Phys.\ Lett.\  B {\bf 302}, 171 (1993)
  [arXiv:gr-qc/9302019].


\bibitem{Destri:2009wn}
  C.~Destri, H.~J.~de Vega and N.~G.~Sanchez,
  arXiv:0906.4102 [astro-ph.CO],
  D.~Cirigliano, H.~J.~de Vega and N.~G.~Sanchez,
  Phys.\ Rev.\  D {\bf 71}, 103518 (2005)
  [arXiv:astro-ph/0412634].



\bibitem{Shafi:1983bd}
  Q.~Shafi and A.~Vilenkin,
  Phys.\ Rev.\ Lett.\  {\bf 52}, 691 (1984).

\bibitem{Pi:1984pv}
  S.~Y.~Pi,
  Phys.\ Rev.\ Lett.\  {\bf 52}, 1725 (1984);
  Q.~Shafi and A.~Vilenkin,
  Phys.\ Rev.\  D {\bf 29}, 1870 (1984).

\bibitem{Lazarides:1984pq}
  G.~Lazarides and Q.~Shafi,
  Phys.\ Lett.\  B {\bf 148}, 35 (1984).

\bibitem{Shafi:2006cs}
  Q.~Shafi and V.~N.~{\c S}eno$\breve{\textrm{g}}$uz,
  Phys.\ Rev.\  D {\bf 73}, 127301 (2006)
  [arXiv:astro-ph/0603830].



\end{thebibliography}
\end{document}